\documentclass[twocolumn,showpacs,preprintnumbers,amsmath,amssymb,superscriptaddress]{revtex4}
\usepackage[bookmarks=false]{hyperref}
\usepackage{amssymb}
\usepackage{amsmath}
\usepackage{graphicx}% Include figure files
\usepackage{dcolumn}% Align table columns on decimal point
\usepackage{bm}% bold math
\usepackage{epstopdf}
\usepackage{times}
\usepackage{color}
\begin{document}
\preprint{}
%
% ###############################################################################################
\title{Exchange bias-like effect in TbFeAl intermetallic induced by atomic disorder}
% ###############################################################################################
%
%\shorttitle{} %Insert here a short version of the title if it exceeds 70 characters
 %Title of paper
% repeat the \author .. \affiliation  etc. as needed
% \email, \thanks, \homepage, \altaffiliation all apply to the current author.
% Explanatory text should go in the []'s, 
% actual e-mail address or url should go in the {}'s for \email and \homepage.
% Please use the appropriate macro for the type of information
% \affiliation command applies to all authors since the last \affiliation command. 
% The \affiliation command should follow the other information.
%% 
%
%
\author{Harikrishnan S. Nair}
\email{h.nair.kris@gmail.com, hsnair@uj.ac.za}
\affiliation{Highly Correlated Matter Research Group, Physics Department, P. O. Box 524, University of Johannesburg, Auckland Park 2006, South Africa}
\author{Andr\'{e} M. Strydom}
\affiliation{Highly Correlated Matter Research Group, Physics Department, P. O. Box 524, University of Johannesburg, Auckland Park 2006, South Africa}
\affiliation{Max Planck Institute for Chemical Physics of Solids (MPICPfS), N\"{o}thnitzerstra{\ss}e 40, 01187 Dresden, Germany}
\begin{abstract}
Exchange bias-like effect observed in the intermetallic compound TbFeAl, which displays a magnetic phase transition at $T^h_c \approx$ 198~K and a second one at $T^l_c \approx$ 154~K, is reported. 
{\em Jump}-like features are observed in the isothermal magnetization, $M (H)$, at 2~K which disappear above 8~K. The field-cooled magnetization isotherms below 10~K show loop-shifts that are reminiscent of exchange bias, also supported by {\em training effect}. Significant coercive field, $H_c \approx$ 1.5~T at 2~K is observed in TbFeAl which, after an initial increase, shows subsequent decrease with temperature. 
The exchange bias field, $H_{eb}$, shows a slight increase and subsequent leveling off with temperature. It is argued that the inherent crystallographic disorder among Fe and Al and the high magnetocrystalline anisotropy related to Tb$^{3+}$ lead to the exchange bias effect. 
TbFeAl is recently reported to show magnetocaloric effect and the present discovery of exchange bias makes this compound a multifunctional one. 
The result obtained on TbFeAl generalizes the observation of exchange bias in crystallographically disordered materials and gives impetus for the search for materials with {\em exchange bias induced by atomic disorder.}
\end{abstract}
\maketitle
\begin{figure}[!t]
\includegraphics[scale=0.70]{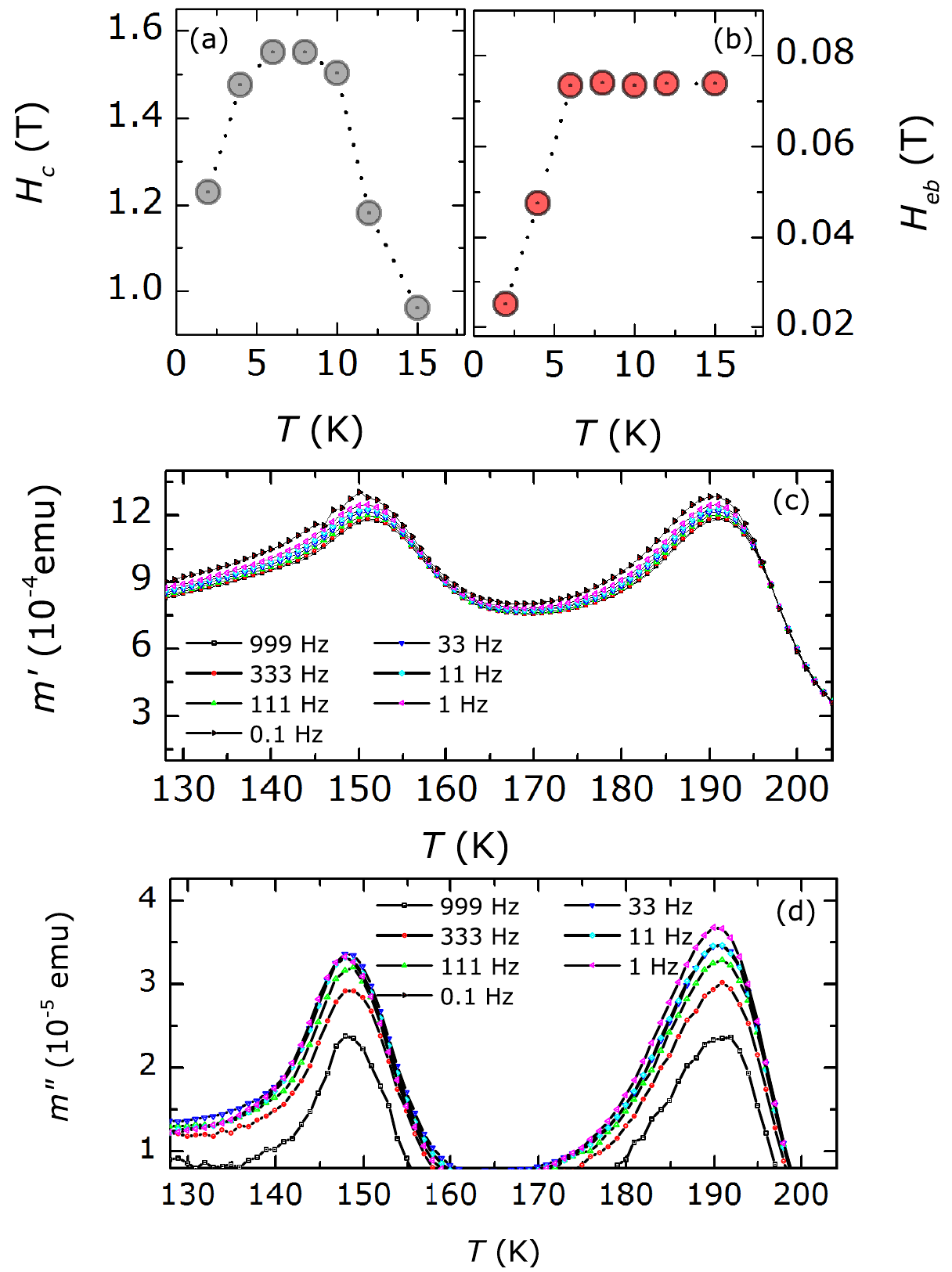}
\caption{\label{fig_eb_acchi} 
The temperature-dependence of coercive field $H_c$ and exchange bias field $H_{eb}$ are shown in (a) and (b) respectively (the error-bars were comparable to the size of data points). 
The real part of the ac susceptibility, $\chi (T)$, is presented for various frequencies 0.1~Hz to 999~Hz in (c) and the imaginary part in (d). 
No significant frequency dispersion is discernible, ruling out the possibility of the presence of canonical spin glass phase. 
}
\end{figure}
\begin{figure}[!t]
\includegraphics[scale=0.40]{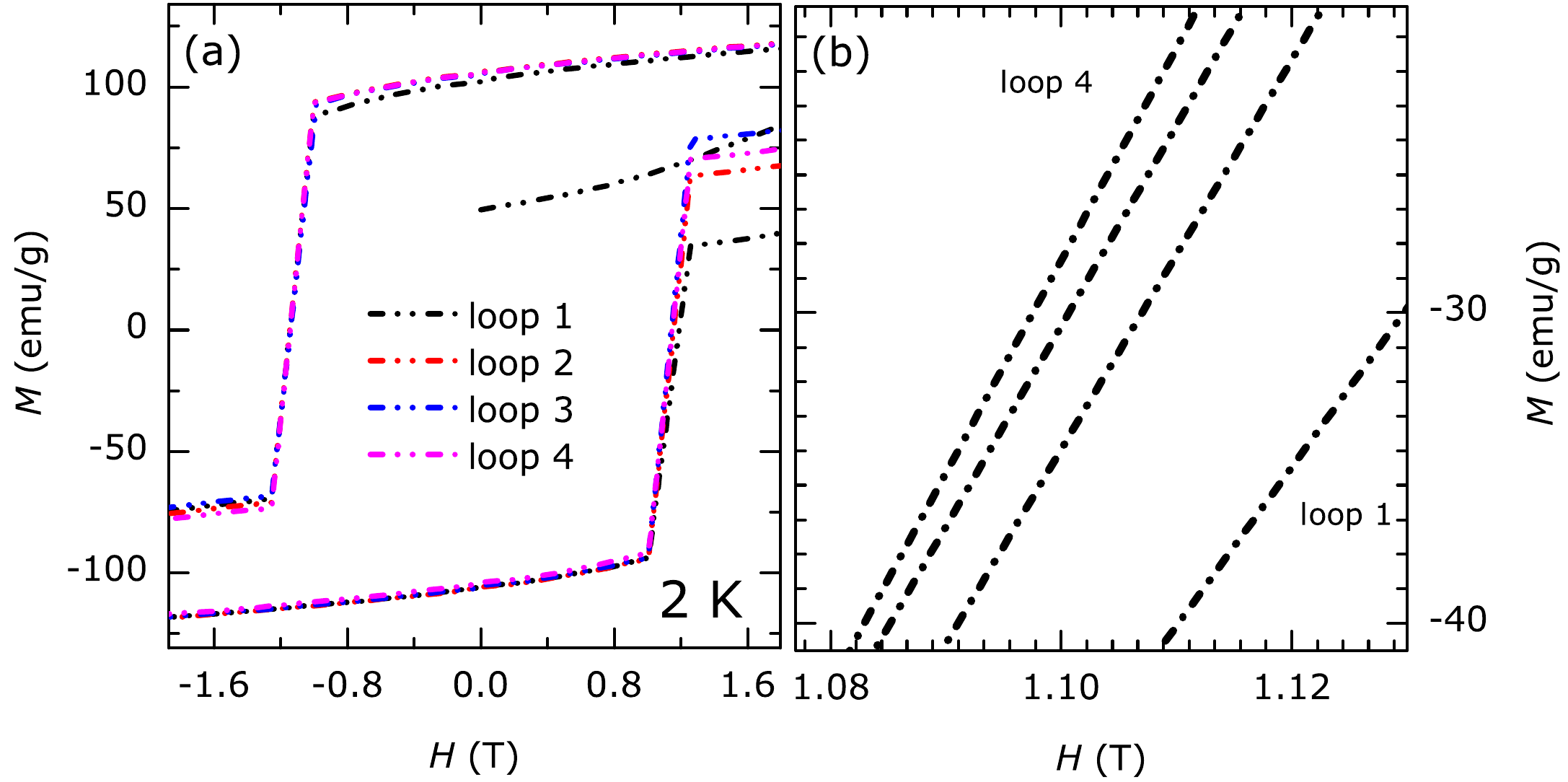}
\caption{\label{fig_training} 
The result of {\em training effect} experiment at 2~K is shown in (a) for 4 loops of hysteresis. A shift of the hysteresis curve with increasing number of loops is visible. In (b), the loops are shown magnified, which confirms the loop-shifts and hence, exchange bias.
}
\end{figure}

%
%
% INTRODUCTION
Reviews on exchange bias\cite{giri_jpcm_23_073201_2011exchange,nogues_1999,kiwi2001exchange,berkowitz1999exchange} in materials point toward the importance of this effect in read-heads in magnetic recording,\cite{zhang_ieee_38_1861_2002magnetic} giant magnetoresistive random access memory devices\cite{tehrani_ieee_91_703_2003magnetoresistive} and in permanent magnets.\cite{sort_apl_79_1142_2001coercivity} 
Interpreted as phenomena occurring at the interface between magnetically ordered microscopic regions, exchange bias interfaces can be ferromagnetic/antiferromagnetic or ferromagnetic/antiferromagnetic/spin-glass.
Mixed magnetic interactions are deemed to be an important ingredient for this effect to occur.
Recent work\cite{morales2015exchange} promotes the importance of ferromagnetic spin structure in exchange bias and explains some of the anomalous features of exchange bias field.
In this Letter we report on the observation of exchange bias-like effect in TbFeAl. 
$R$FeAl ($R$ = rare earth) were first investigated by Oesterreicher.\cite{oesterreicher1971magnetic,oesterreicher1973structural,oesterreicher1977magnetic}
Only the heavier rare earths were observed to form the stable MgZn$_2$-type structure. 
TbFeAl is recently identified as a magnetocaloric\cite{kavstil2014magnetic} (albeit, a weak effect) which becomes a possibel "multifunctional" compound with the observation of exchange bias.
\\
%
%
%-------------Figure--------------------------------------------------%
\begin{figure}[!t]
\includegraphics[scale=0.09]{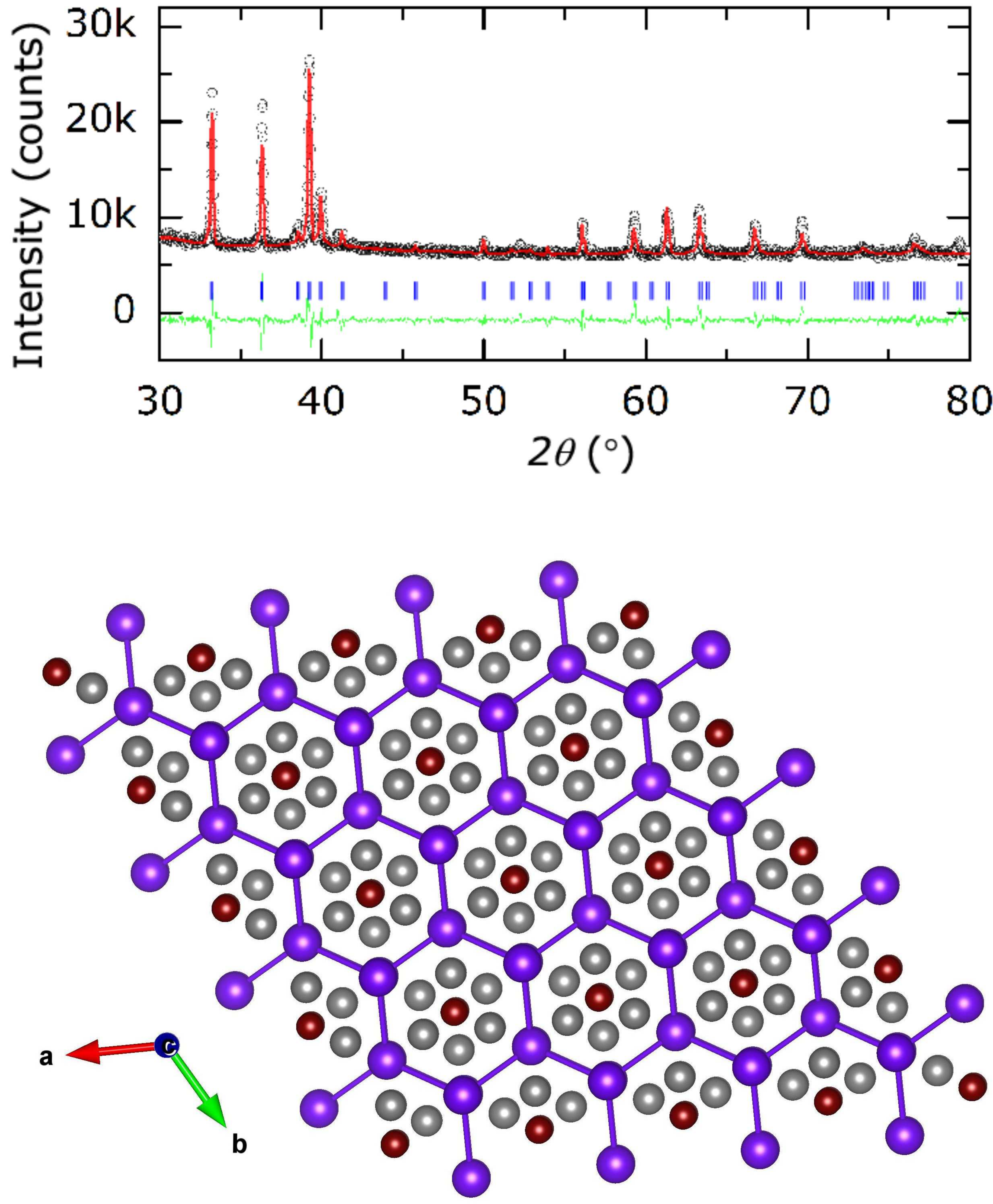}
\caption{\label{fig_xrd} 
{\bf Top:} The powder x ray diffraction pattern of TbFeAl along with Rietveld refinement. 
The black circles are the experimentally observed intensity, the red solid line is the calculated intensity assuming $P6_3/mmc$ space group. 
The difference curve is shown as green solid line and the allowed Bragg peaks as vertical bars. 
{\bf Bottom:} The hexagonal framework of TbFeAl shown as a projection on to the $ab$-plane.
}
\end{figure}
%-------------Figure--------------------------------------------------%
%
%
\indent
As noted above, TbFeAl crystallizes in hexagonal MgZn$_2$ type crystal structure with $P6_3/mmc$ space group. 
Tb occupies $4f$ Wyckoff position while Fe and Al are situated at $2a$ and $6h$ sites respectively.\cite{oesterreicher1973structural,oesterreicher1977magnetic}
Mixed occupation of Fe and Al is possible in this structure.
Magnetically, TbFeAl is a ferrimagnet with a transition temperature of 195~K.\cite{kavstil2014magnetic} 
A saturation-like effect of magnetization which displayed an $S$-curve was observed\cite{oesterreicher1971magnetic} which was described as a consequence of partial chemical disorder of Fe and Al. 
Significantly high magnetocrystalline anisotropy leading to the formation of thin domain walls which are pinned to defects influences the magnetization, and to some extent, is the reason for the magnetocaloric effect.\cite{kavstil2014magnetic}
\\
\indent
Partial or total crystallographic disorder is a favourable ingredient for the observation of exchange bias. 
Granular nanoparticles in a structurally and magnetically disordered matrix\cite{fiorani2007exchange} or interacting magnetic defects embedded in an antiferromagnetic matrix with high degree of disorder\cite{gruyters2009interacting} are example systems where {\em disorder} brings about exchange bias effect.
In the oxide Y$_2$CoMnO$_6$, below 8~K prominent {\em steps} in magnetization and significant coercive field of $\approx$ 2~T were observed.\cite{nair2015antisite} 
Martensitic-like growth of ferromagnetic domains, formed as a result of {\em antisite} disorder, was postulated as the reason for exchange bias in Y$_2$CoMnO$_6$. 
Motivated by the prospect of obtaining a general feature of {\em exchange bias induced by disorder}, we have extended our research to a disordered intermetallic -- TbFeAl.
\\ 
\indent 
The polycrystalline sample used in this study was prepared using arc melting method. 
The elements Tb, Fe and Al (all $4N$ purity) were melted in the water-cooled Cu hearth of an Edmund Buehler furnace in Argon atmosphere. 
The once-melted buttons were remelted four or five times to ensure homogeneity. 
Post melting, powder x ray diffractograms were recorded on pulverized samples in a Rigaku SmartLab x ray diffractometer which used Cu K-$\alpha$ radiation. 
Magnetic properties were recorded using a Magnetic Property Measurement System from Quantum Design Inc., San Diego. Magnetization as a function of temperature in the range 2 - 350~K in both zero field-cooled (ZFC) and field-cooled (FC) protocols as well as magnetic field in the range 0 - 7~T and ac susceptibility measurements were performed.
\\
\indent
The experimentally obtained powder x ray diffractogram is presented in Fig~\ref{fig_xrd} as black circles. 
The observed peaks could be indexed in hexagonal space group $P6_3/mmc$ (MgZn$_2$ type). 
FullProf suite of programs\cite{fullprof} was used to perform Rietveld analysis\cite{rietveld} of the x ray data which yielded $a$~({\AA}) = 5.3975(4)  and $c$~({\AA}) = 8.7526(3). 
The results of the Rietveld refinement are presented in Fig~\ref{fig_xrd}.
In the bottom panel of Fig~\ref{fig_xrd}, the hexagonal structure of TbFeAl is shown as a projection on to the $ab$ plane.
\\
%
%
%---------------------Figure------------------------------------------------%
\begin{figure}[!t]
\includegraphics[scale=0.48]{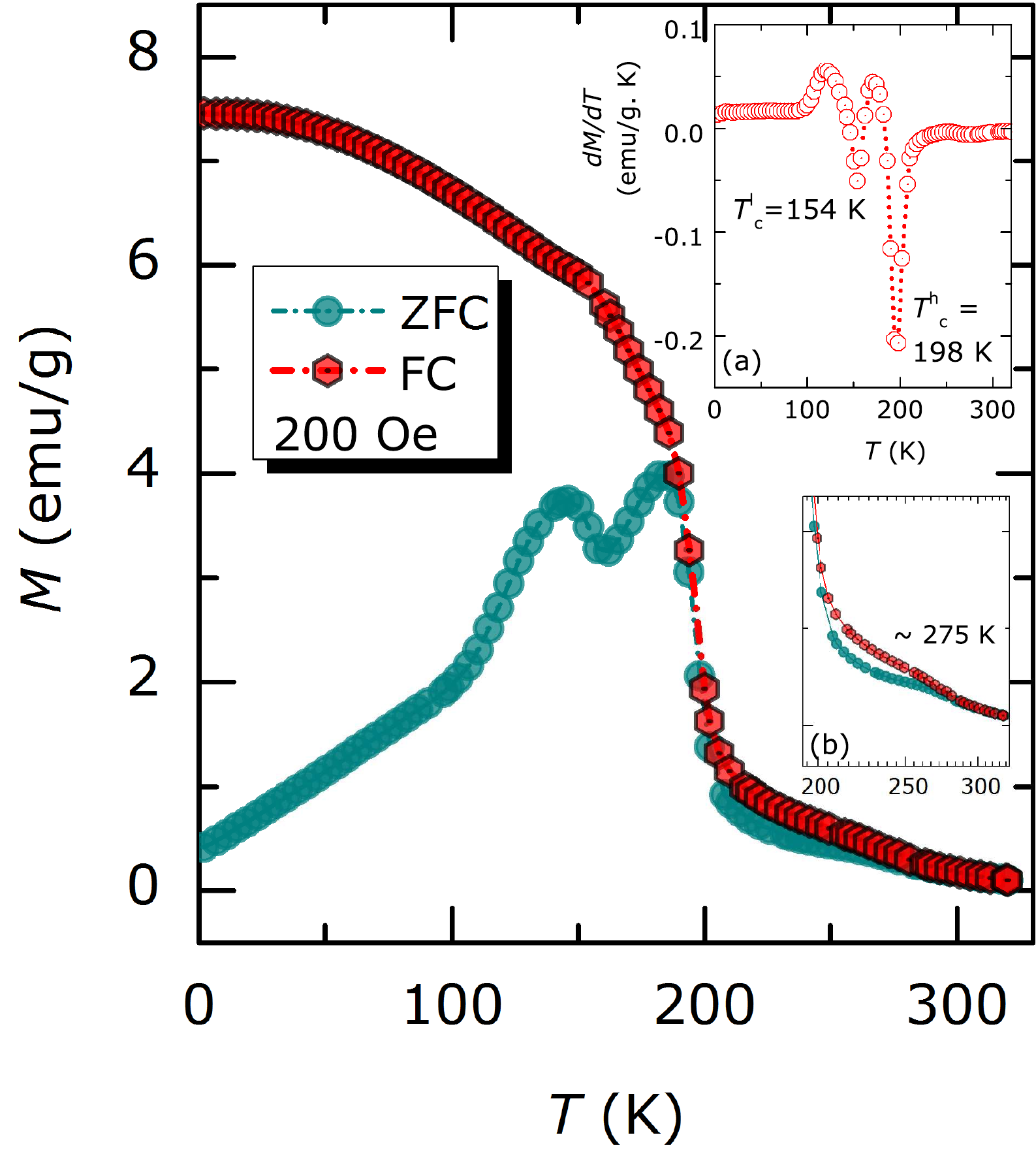}
\caption{\label{fig_mt} 
Magnetization curves (ZFC and FC) of TbFeAl obtained at 200~Oe. 
Two magnetic transitions at $T^h_c \approx 198$~K and $T^l_c \approx 154$~K are observed. 
Inset (a) shows the derivative plot, $dM/dT$, used to determine the transition temperatures. 
Inset (b) magnifies the region close to 250~K ($> T^h_c$) where a "loop"-like feature is seen.
}
\end{figure}
%---------------------Figure------------------------------------------------%
%
%
\indent
The magnetic phase transitions reported in the literature for TbFeAl\cite{kavstil2014magnetic,klimczak2010magnetocaloric,li2014study} are reproduced in the magnetization curve, $M (T)$, obtained at 200~Oe which is presented in Fig~\ref{fig_mt}. 
A bifurcation between the ZFC and FC arms is observed at $\approx$ 275~K (see inset (b)) followed by two {\em humps} at 186~K and at 145~K. The transition temperatures are determined by taking the derivative of $M (T)$ and is plotted as $dM/dT$ in the inset (a). 
From this, $T^h_c \approx$ 198~K and $T^l_c \approx$ 154~K are determined. 
In the literature, the second transition is attributed to the existence of two crystallographic regions in the sample with different occupation of Fe and Al on the sites $2a$ and $6h$. 
The FC arm of $M (T)$ suggests that upon application of magnetic field, ferromagnetic like enhancement of magnetization results. 
In the inset (b) of Fig~\ref{fig_mt}, the high temperature region of $M (T)$ for $T \geq T^h_c$ is shown magnified to highlight a "loop"-like structure. 
It is to be noted that no significant linear region is observed up to 350~K and hence a description of magnetic susceptibility following ideal Curie-Weiss formalism does not hold. 
The "loop"-like structure in $M (T)$ at high temperature might suggest the presence of magnetic correlations extending above $T^h_c$.
\\
%
%
%-----------------------Figure----------------------------------------%
\begin{figure}[!b]
\includegraphics[scale=0.41]{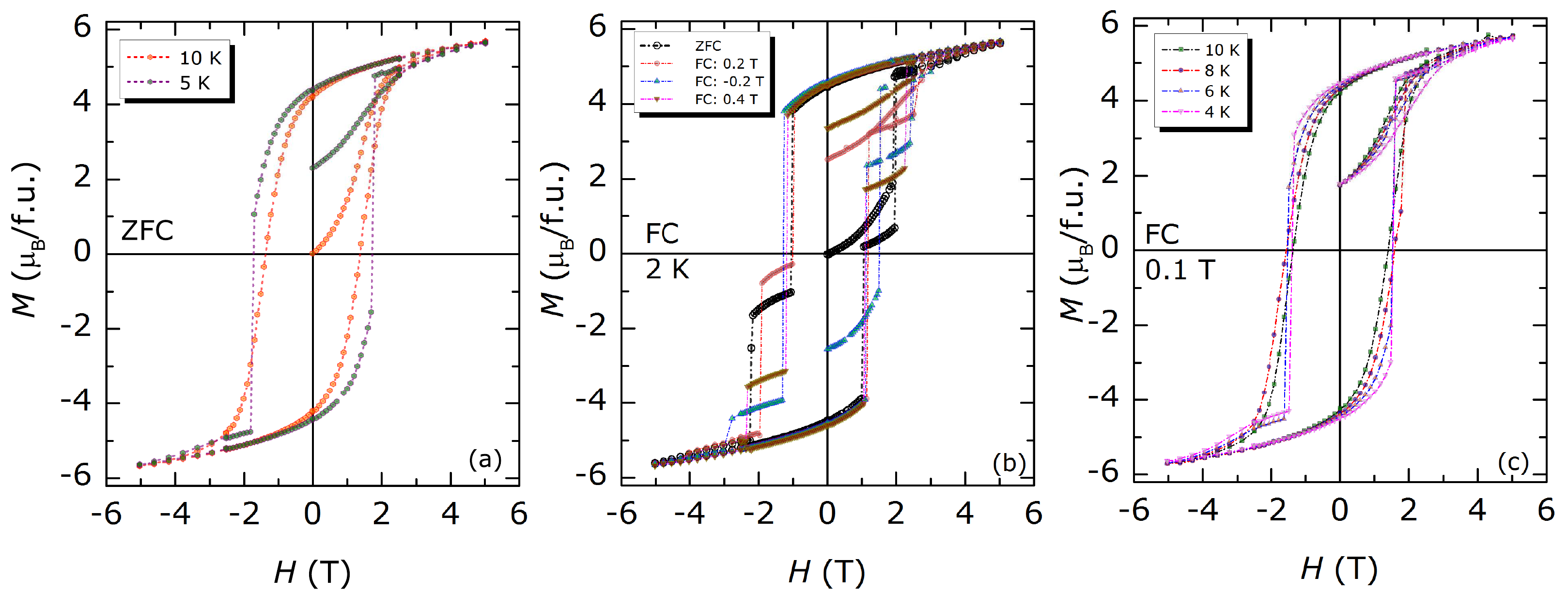}
\caption{\label{fig_mh_1} 
(a) Isothermal magnetization curves, $M (H)$, in zero field-cooled mode for TbFeAl measured at 5~K and 10~K. 
At 10~K, no {\em jumps} in magnetization are visible while at 5~K they are present. 
Significant coercive field is observed in both the cases and the saturation magnetization attains $\approx$ 5.8~$\mu_\mathrm{B}/$f.u. 
(b) The magnetization isotherms at 2~K obtained under different values of field cooling. 
The ZFC curve is also plotted for a comparison. (c) The magnetization isotherms at 4, 6, 8 and 10~K after field cooling in 1000~Oe from 300~K. 
The field-cooled curves display clear evidence for loop-shifts reminiscent of exchange bias effect.
}
\end{figure}
%-----------------------Figure----------------------------------------%
%
\indent 
The ferromagnetic feature of TbFeAl is evident from the isothermal magnetization curve, $M (H)$, at 5~K and 10~K presented in Fig~\ref{fig_mh_1} (a).
Significant coercive field ($H_c \approx$ 1.5~T) and magnetic saturation-like effects are observed at 2~K. 
The $M (H)$ curves in Fig~\ref{fig_mh_1} (a) are comparable to the $S$-curve reported in TbFeAl.\cite{oesterreicher1971magnetic,oesterreicher1977magnetic,kavstil2014magnetic}
Interestingly, sharp {\em jumps} of magnetization are observed in the zero field-cooled magnetization isotherm at 5~K, see Fig~\ref{fig_mh_1} (a). 
As the temperature is reduced, the sharp {\em jumps} in $M (H)$ are enhanced. 
The $M (H)$ curves in ZFC mode at 2~K presented in Fig~\ref{fig_mh_1} (b) display the {\em jumps}. 
In (b), the magnetization isotherms at 2~K obtained in field cooled mode under different applied fields are presented. 
With the application of magnetic field, the hysteresis loops shift from the ZFC location though the displacement is not highly systematic. 
This observation is reminiscent of exchange bias. Similar features are observed in (c) where isotherms at different temperatures are presented, all obtained after field-cooling in 1000~Oe from 300~K to
the temperature of measurement. 
As the temperature is increased, the width of the hysteresis loop decreases and the {\em jumps} disappear. 
At 2~K the maximum magnetization attained by the application of 5~T is $\approx$ 5.8~$\mu_\mathrm{B}/$f.u., which is significantly reduced from the free-ion moment of Tb$^{3+}$
which is 9~$\mu_\mathrm{B}$.\cite{oesterreicher1971magnetic,oesterreicher1977magnetic}
\\
\indent
The exchange bias field $H_{eb}$ = $(H_+ + H_-)/2$ (where $H_{\pm}$ are the positive and negative intercepts of the magnetization curve with the field axis) and the coercive field, $H_c$, were estimated from the data in Fig~\ref{fig_mh_1} (b,c). 
Figure~\ref{fig_eb_acchi} (a,b) present the evolution of $H_c$ and $H_{eb}$ as a function of temperature. 
Though TbFeAl displays significant $H_c$ (about 1.5~T at 6~K), it is lower than that of Y$_2$CoMnO$_6$ which has similar domain-related structure.\cite{nair2015antisite} 
Generally, a monotonous decrease of $H_c$ with increasing temperature is favoured. 
However, disordered granular systems are reported to show a variation of $H_c$ similar to what has been observed for TbFeAl.\cite{fiorani2007exchange} 
The anomalous temperature dependence of $H_{eb}$ could be related to the recent work on the role of ferromagnetic layers or domains in exchange bias\cite{morales2015exchange} where, contrary to the conventional case, an increase of $H_{eb}$ with temperature is explained.
The atomic disorder in TbFeAl could be held responsible for such a behaviour. 
Interestingly, though the $H_c$ of TbFeAl shows an initial increase and subsequent decrease with temperature, $H_{eb}$ increases first and attains a near-constant value up to 15~K.
\\
\indent
Exchange bias effect results from interfaces between ferromagnetic, antiferromagnetic or spin-glass regions. 
In order to probe the presence of spin glass in TbFeAl, ac susceptibility measurements were carried out at frequencies ranging from 0.1~Hz to 999~Hz. 
The results are presented in Fig~\ref{fig_eb_acchi} (c). The susceptibility peaks at $T^l_c$ and $T^h_c$ are observed to show no frequency dependence other than weak damping. 
The peak positions were determined by taking the derivative $d\chi' (f, T)/dT$. 
It is then clear that the disorder in this material only pertains to structural aspects. 
It was unable to quantify the degree of disorder related to the mixed-occupancy between Fe and Al from the x ray data. 
However, with the introduction of mixed-occupancy, an improvement in the goodness-of-fit of refinement was observed. 
Presence of nano magnetic domains in $R$FeAl were experimentally observed in the case of TmFeAl\cite{mulders2000observation} where the crystallographic disorder and the high magnetocrystalline 
anisotropy of Tm were the reason. 
By comparison, the experimental data presented here for TbFeAl suggests a similar scenario. 
Finally, the loop-shifts observed in {\em training effect} experiment presented in Fig~\ref{fig_training} confirms the exchange bias in TbFeAl.
Hysteresis loops were measured at 2~K for 4 continuous loops after field cooling the sample using 1000~Oe.
It can be seen that the hysteresis curves begin to shift with increasing number of loops.
For clarity, only a part of the hysteresis is shown in (a).
A fully magnified view of the loop-shift is provided in (b).
% 
%
%-----------------------------------Figure----------------------------------------%
%-----------------------------------Figure----------------------------------------%
%
%
We now employ a similar analysis as was employed in the case of Y$_2$CoMnO$_6$ which also exhibited domain-related effects.\cite{nair2015antisite}
The scenario of pinning of domain walls with associated lattice strain has been modeled in phase-separated manganites.\cite{niebieskikwiat_jpcm_24_436001_2012pinning}
Ferromagnetic clusters in structurally disordered materials present strain fields that can pin the domain walls. 
The formalism of pinning of elastic objects described above is similar to the pinning of vortices in high-$T_c$ superconductors.\cite{larkin_31_784_1970effect} 
According to the model developed by Larkin, the surface tension of a typical domain wall is approximated as $\epsilon_t \sim 4\sqrt{\frac{JS^2K}{a}}$ where $J$ is the exchange constant between nearest neighbour spins of magnitude $S$, $K$ is the magnetocrystalline anisotropy energy per unit volume and $a$ is the distance between nearest neighbour spins.  
The $J$ value for TbFeAl can be approximated by the value $J_\mathrm{Tm-Tm}\approx$ 1~K for TmFeAl.\cite{mulders1998complex}
The spin value for Tb$^{3+}$ is $S$ = 3 and a value of $K \approx$ 10$^3$ J/m$^3$ is adopted from the typical values for TmFeAl.\cite{mulders1996unique}
From the structural refinement using x ray data, the nearest neighbour distance $a$ for TbFeAl (Tb-Tb distance) is obtained as 0.343~nm. With these values, an $\epsilon_t \sim$ 2.1 $\times$10$^{-4}$~J/m$^2$. 
The amplitude of surface excitations follows the relation $\xi \sim \sqrt{\frac{U}{2\epsilon_t}}$. 
The ac susceptibility data of TbFeAl above the $T^h_c$ could be roughly described by a thermally activated behaviour, $\chi(T)$ = $\chi_0$ + $\chi_1~ \mathrm{exp}(-U/k_BT)$ yielding a value, $U \sim$ 83~meV.  
Substituting these values, $\xi \approx$ 5.5~nm. 
\\
%
%
%-----------------------------------Figure----------------------------------------%
%-----------------------------------Figure----------------------------------------%
%
%
\indent
Among $R$FeAl ($R$ = rare earth) compounds, TmFeAl is reported to display unique magnetic properties.\cite{mulders1996unique,mulders2000observation,mulders1998complex}
The formation of nano-magnetic domains in TmFeAl was discovered through M\"{o}{\ss}bauer spectroscopy.\cite{mulders1996unique}
The magnetic structure of TmFeAl describes ferromagnetic sublattice of Tm moments aligned along the $c$-axis of the hexagonal  cell. 
Fe and Al are arranged disordered in the unit cell and because of the low moment of Fe ($\sim$ 0.52~$\mu_B$), it is not easily detected in neutron diffraction studies. 
However, at low temperature a development of ferromagnetic short-range order of the size of 1~nm is inferred. 
Interestingly, this size corresponds to the size of nano magnetic domains observed through neutron depolarization investigations.\cite{mulders2000observation}
The disorder in the Fe-Al sublattice combined with the high magnetocrystalline anisotropy of Tm moments were attributed as the reason for the development of nano domains. 
The anisotropy field of Tm is about 100~T\cite{mulders1998complex} and hence, can prevent the nano domains from ordering under applied fields.
\\
\indent 
Exchange bias in ferromagnetic TbFeAl with significant coercive field is experimentally demonstrated. 
$H_c \approx$ 1.5~T is observed at 2~K. 
Magnetic saturation is observed in isothermal magnetization curves below 5~K however, not reaching the full ferromagnetic moment of Tb$^{3+}$. 
The crystallographic disorder among Fe-Al sublattice and the strong magnetocrystalline anisotropy of Tb$^{3+}$ are argued to be the reason for observed domain effects that lead to exchange bias. 
The results presented in this work attains a general feature following the extension from exchange bias effect observed in double perovskite Y$_2$CoMnO$_6$ which was driven by the theme {\em exchange bias induced by atomic disorder.}
\\
\indent
HSN acknowledges FRC/URC for the grant of a postdoctoral fellowship and AMS thanks the SA NRF (93549) and UJ URC/FRC for financial assistance.
\\ \\
$\#$Present address: Department of Physics, Colorado State University, Fort Collins, CO 80523
%
%
% % % % % % % % % % % % % % % % % % % % % % % % % % % % 
%\bibliography{EB_TbFeAl}
%\bibliographystyle{apsrev}
%% % % % % % % % % % % % % % % % % % % % % % % % % % % % 
%

%
\end{document}